\UseRawInputEncoding
\documentclass{aastex631}

\usepackage{orcidlink}
\usepackage{hyperref}
\usepackage{booktabs}
\usepackage[T1]{fontenc}

\begin{document}


\title{Pinching of ICME Flux Rope: Unprecedented Multipoint Observations of Internal Magnetic Reconnection during Gannon's Superstorm}

\author[0009-0005-7637-8346]{Shibotosh Biswas}
\affiliation{Space Physics Laboratory, Vikram Sarabhai Space Centre, Indian Space Research Organization,  Trivandrum,  Kerala 695022, India}
\altaffiliation{Department of Physics, University of Kerala, Trivandrum, India}

\author[0000-0003-4281-1744]{Ankush Bhaskar}
\affiliation{Space Physics Laboratory, Vikram Sarabhai Space Centre, Indian Space Research Organization, Trivandrum,  Kerala 695022, India}

\author[0000-0002-4704-6706]{Anil Raghav}
\affiliation{Department of Physics, University of Mumbai, Vidyanagari, Santacruz (E), Mumbai 400098, India}

\author[0009-0007-1070-3649]{Ajay Kumar}
\affiliation{Department of Physics, University of Mumbai, Vidyanagari, Santacruz (E), Mumbai 400098, India}

\author[0000-0002-9506-6875]{Kalpesh Ghag}
\affiliation{Department of Physics, University of Mumbai, Vidyanagari, Santacruz (E), Mumbai 400098, India}

\author[0000-0002-0116-829X]{Smitha V. Thampi}
\affiliation{Space Physics Laboratory, Vikram Sarabhai Space Centre, Indian Space Research Organization, Trivandrum, Kerala 695022, India}

\author[0000-0002-1470-8443]{Vipin K Yadav}
\affiliation{Space Physics Laboratory, Vikram Sarabhai Space Centre, Indian Space Research Organization, Trivandrum, Kerala 695022, India}

\begin{abstract}
The extreme solar storm of May 10, 2024, during the 25th solar cycle, which recorded a symmetric H component index (Sym-H) reaching -500 nT, was the strongest since the 2003 Halloween storm. This event offered a unique opportunity for unprecedented multipoint observation of the complex interaction of  Interplanetary Coronal Mass Ejections (ICME) from different vantage points. Utilizing NASA's Wind, ACE, DSCOVR, THEMIS-C, STEREO-A, MMS, and ISRO's recently launched Aditya-L1 spacecraft, we comprehensively investigated the spatio-temporal variations in interplanetary plasma and magnetic field parameters. Our study reveals large-scale quasi-steady magnetic reconnection within the interior of the ICME flux rope, possibly triggered by interactions between multiple ICMEs. A current sheet (CS) forms within the flux rope, enabling internal magnetic reconnection between concentric magnetic surfaces, which leads to a sharp reversal of the IMF $B_{y}$ component, as observed at the L1 point. Concurrently, reconnection exhaust and enhanced electron and ion fluxes were detected with the CS, extending over 200 $R_{E}$ (1.3 million km) along the GSE-y direction. This finding sheds new light on the role of internal reconnection in ICME evolution, highlighting its pivotal role in modifying the morphology of the ICME magnetic structure and exerting severe space weather effects on Earth.

\end{abstract}


\section{Introduction}\label{sec:intro}
In our increasingly technology-dependent world, extreme space weather has emerged as a critical concern. Solar eruptions, particularly coronal mass ejections (CMEs), modulate the heliospheric plasma and are the primary drivers of severe space weather events \citep{kilpua2017coronal}, disrupting satellite operations, communication networks, GPS systems, and power grids, posing substantial risks to infrastructure and human activities \citep{hapgood2017space, Hayakawa_2020}. A crucial challenge in space weather prediction is accurately determining the arrival time of interplanetary CMEs (ICMEs) at the Earth  \citep{riley2018extreme}, but equally important is understanding their magnetic topology. The orientation and internal structure of an ICME magnetic field critically influence solar Wind-magnetosphere energy transfer, affecting the severity of geomagnetic storms \citep{teng2024unexpected, gonzalez2011interplanetary}. Thus, precise predictions of both the ICME arrival and morphology are vital for mitigating the impact of space weather.\\

One such extreme solar and geomagnetic storm occurred from May 10 to 12, 2024, known as Gannon's or Mother's Day Storm. The Earth's magnetopause (a boundary between solar and magnetospheric plasma) was compressed below 5 Earth radii (1 $R_{E}$ = 6371 km) \citep{fu2025compression}, exposing geostationary satellites to harsh solar wind conditions. Although geomagnetic disturbances on Earth are primarily driven by the southward component of the interplanetary magnetic field (IMF-$B_{Z}$) \citep{gonzalez1994geomagnetic,ashna2024solar}, studies have established the critical impact of IMF-$B_{y}$ also \citep{holappa2020explicit, holappa2021explicit, chakrabarty2017role}. A sudden change in the IMF-$B_{y}$ and IMF-$B_{Z}$ orientation alters the coupling between solar wind and magnetosphere. These transient events affect the efficiency of energy transfer from the solar wind to the ionosphere through the magnetosphere \citep{vichare2024manifestations,ohtani2025ground, zhang2025double, venugopal2025electrodynamic}, which modulates auroral and even low-latitude ionospheric activities significantly. However, the underlying cause of this drastic change in IMF orientation and morphology remained unclear, prompting a detailed investigation into its origin. \\   


Multiple CMEs from the Sun arriving on May 10 \& 11 were responsible for the intense geomagnetic storm observed on Earth \citep{kwak2024observational}. CMEs are plasma bursts with magnetic fields that release an enormous amount of energy into interplanetary space from the solar corona \citep{gosling1974mass, st2000properties}. The first ICME observations in the 1970s revealed loop- or bubble-like structures associated with closed magnetic fields, often described as magnetic flux ropes (MFRs) \citep{Hirshberg1970Observation, Gosling1973Anomalously, Palmer1978Bidirectional}. The trajectory of the observing spacecraft decides the inferred structure of the ICME. If the observation displays: (1) enhanced magnetic field compared to surroundings ($>10 \, nT$ ); (2) smooth rotation of magnetic field components over a large angle; and (3) reduced proton temperature and plasma beta (ratio between thermal and magnetic pressure of plasma) \citep{kilpua2017coronal, klein1982interplanetary}, the ICME can be classified as a Magnetic Cloud (MC)\citep{burlaga1981magnetic}. ICME flux ropes might have high plasma beta; however, only those flux ropes with the above properties are denominated magnetic clouds. These ICMEs generate a shock ahead of them due to their relatively higher speed compared to the communication speed in the solar wind. The region between the shock front and the MC is filled with a turbulent and compressed plasma medium with a fluctuating magnetic field, high plasma density, and temperature, known as the ICME sheath. Comprehensive reviews of ICMEs and their solar-terrestrial interactions are available in the literature \citep{gopalswamy2016history,kilpua2017coronal, webb2012coronal}.  \\ 

ICMEs undergo dynamic transformations, such as twisting, pinching, kinking, and magnetic reconnection, throughout their evolution, from initiation to propagation, while interacting with the ambient solar wind (\cite{manchester2005coronal,  niembro2019numerical, pal2020flux, wang2016propagation} and references therein) and significantly alters their morphology, often flattening them into a \lq{}Pancake\rq{} \citep{manchester2004three, raghav2019cause}. For example, twisting magnetic field lines can cause kinking and pinching within the MC, potentially fragmenting it into multiple large-scale plasmoids \citep{khabarova2015small, khabarova2016small}. Within these compressed flux ropes, reconnection is expected to occur frequently due to variations in magnetic field orientations \citep{khabarova2021current}. In short, magnetic reconnection is a fundamental plasma process in the universe in which magnetic energy is efficiently converted into kinetic energy, heat, and particle acceleration \citep{gonzalez2016magnetic} and could play a significant role in ICME evolution in the heliosphere.\\

A considerable number of studies have been conducted to date for this extreme solar event of May 2024 from its origin, propagation, and effect on geospace \citep{kruparova2024unveiling, kwak2024observational, vichare2024manifestations, ohtani2025ground, venugopal2025electrodynamic}. This article presents a plausible mechanism for the abrupt change in IMF components around 22:05 UT, as observed by multiple satellites near the Sun-Earth L1 point. Fleet of space observatories during this unique event provided unprecedented in-situ observations from different vantage points using a total of seven spacecraft: NASA's ACE, Wind, DSCOVR, STEREO-A, ARTEMIS-P2, MMS, and recently launched Indian Space Research Organization's first observatory-class solar mission: Aditya-L1, revealing that the reconnection was triggered in the interior of an ICME MC, possibly due to the compression by adjacent ICME MC and quasisteady throughout the observation time. The prominent electron and ion flux enhancements, along with the suprathermal electron pitch angle distribution from the Wind spacecraft, complement the inference. Moreover, we attempted to provide an underlying physical mechanism of the occurrence of magnetic reconnection, utilizing multispacecraft observations. In the following three sections, data, methods, detailed observations, and in-depth discussions are presented. 

\begin{figure*}[htbp]
\centering
\includegraphics[width=0.9\textwidth]{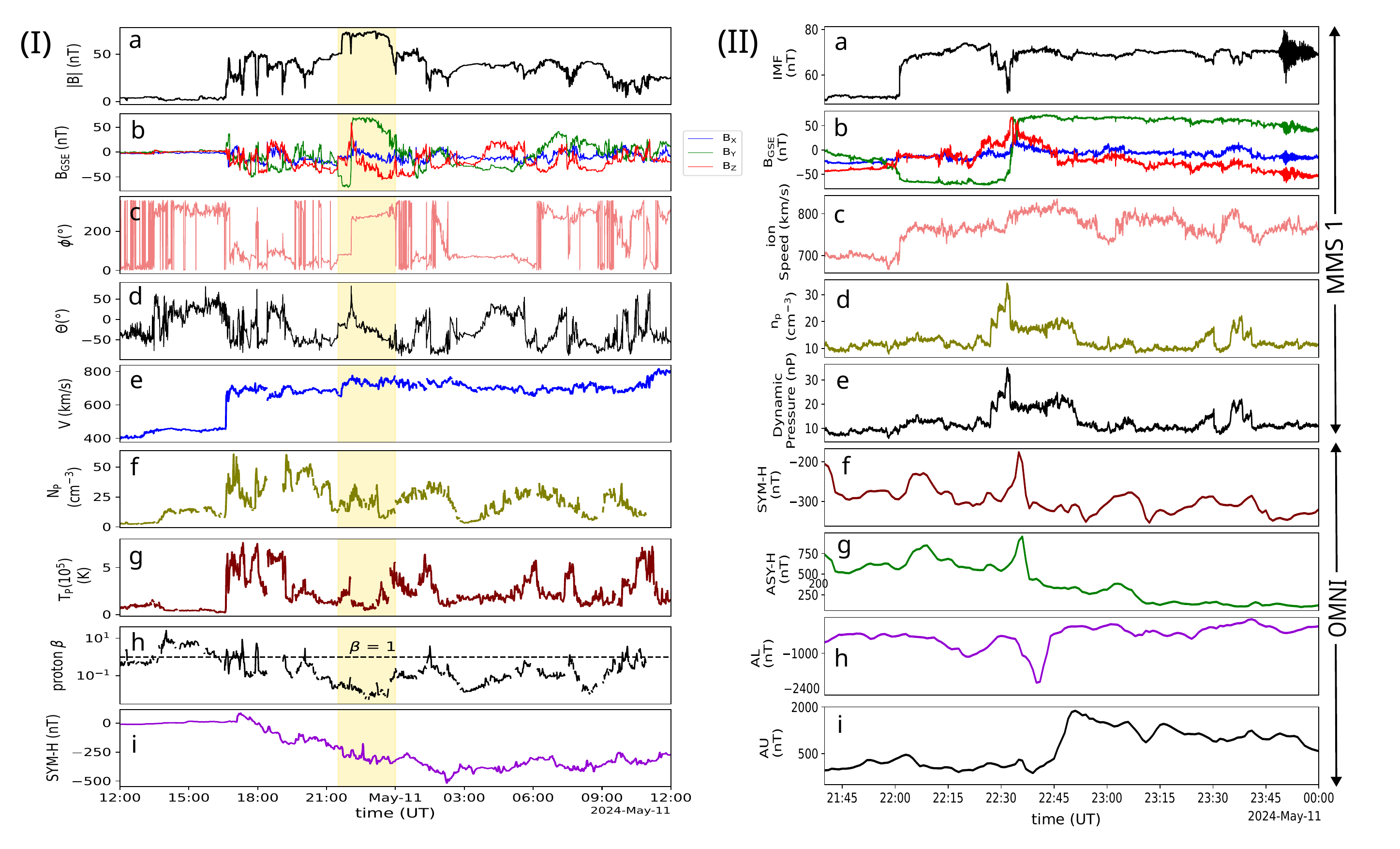}
\caption{Panel(I) displays the observations recorded by the ACE spacecraft from May 10, 12:00 to 23:59:59 UT. Subplanels a to i represent IMF intensity, IMF components, IMF azimuth, elevation, solar wind bulk velocity, proton density, temperature, plasma beta (proton), and Sym-H. The yellow-shaded region marks the region of interest. Panel(II) shows MMS1 and OMNI data of the highlighted region of panel(I) with additional information about dynamic pressure, ASY-H, AL, and AU indices related to geomagnetic storms.}
\label{figure:summary}
\end{figure*}

\section{Data and Methods}
\label{sec:data}

This study uses in situ measurements from multiple upstream solar wind vantage points. From NASA's Advanced Composition Explorer (ACE), we use Level-2 data: 16-second cadence magnetic field measurements \citep{smith1998ace} and 64-second cadence plasma parameters from SWEPAM \citep{mccomas1998solar}. Magnetic field data from ISRO's Aditya-L1 fluxgate magnetometer \citep{Yadav2025} are included. Wind mission observations from the Three-Dimensional Plasma and Energetic Particle Investigation \citep{lin1995three} and Magnetic Field Instrument(3-sec) \citep{lepping1995wind} are used for plasma moments, particle omnidirectional flux, suprathermal electron pitch angle distributions (PAD), and magnetic fields, respectively. NASA's DSCOVR magnetic field(1-sec) and plasma moments are obtained from the PlasMag suite \citep{burt2012deep}. THEMIS fluxgate magnetometer data \citep{angelopoulos2008themis,auster2008themis} are also incorporated. Additionally, Level-1 magnetic field measurements (8 Hz) from the STEREO-Ahead spacecraft are used, obtained from the IMPACT suite's magnetic field experiment \citep{acuna2008stereo,kaiser2008stereo}. For near Earth's magnetopause observation, we have utilized MMS1 fluxgate magnetometer\citep{russell2016magnetospheric} survey data (8 samples/s), Fast Plasma Instrument (FPI)\citep {pollock2016fast}, Dual ion Spectrometer fast mode data for plasma moments, and OMNI data for geomagnetic indices. \\
In addition to plasma and magnetic field data analysis, Minimum Variance Analysis (MVA) and 2D-hodograms of the magnetic field (see Section~\ref{sec:lmn}) are used to infer magnetic flux rope geometry and orientation (see Appendix section~\ref{sec:MVA} for more details). 

\section{Observations and Analysis}\label{sec:Observations}

\subsection{Overview of the Event}
On May 10, 2024, Earth experienced a severe G5-class geomagnetic storm \citep{hayakawa2025solar, Thampi_2025}, triggered by multiple coronal mass ejections (CMEs). Figure~\ref{figure:summary} panel(I) highlights NASA's Advanced Composition Analyser (ACE) solar wind data in the Geocentric Solar Ecliptic (GSE) frame. At 16:37 UT, the ICME shock arrival is marked by a sudden increase in interplanetary magnetic field (IMF) to 50 nT, solar wind velocity jump to 750 km/s, plasma density rising above 50 particles per $cm^3$, and proton temperature exceeding 0.5 million Kelvin. The subsequent ICME sheath featured turbulent magnetic fields, elevated density, and temperature, with possible interaction regions. In a recent study, \cite{weiler2025first} reported preceding CMEs before 20:00 UT using ACE and STEREO-A data. This study focuses on 21:40 to 23:59 UT on May 10 (highlighted in yellow, in figure~\ref{figure:summary}), showing constant $|\vec{B}|$ of 75 nT, followed by a sharp decrease at 22:04 UT, an abrupt rotation of the $B_{y}$ from -70 nT to +75 nT, and $B_{z}$ revarsal. The decrease in plasma density and temperature, and plasma $\beta$ (the ratio of thermal to magnetic pressure), suggests the highlighted region corresponds to an ICME MC. Figure~\ref{figure:summary} panel(II)h and i reports increased auroral Eastward (AL index) and Westward currents (AU index) at that time, indicating enhanced coupling between solar wind-magnetosphere-ionosphere. The enhancement in SYM-H(e) and ASY-H(f) at 22:30 UT suggests a sudden amplification of the magnetopause current by a few hundred nT, due to localized high density and solar wind dynamic pressure.  \\

\subsection{Multipoint signatures of magnetic reconnection}
\begin{figure*}[htbp]
\centering
\includegraphics[width=0.99\textwidth]{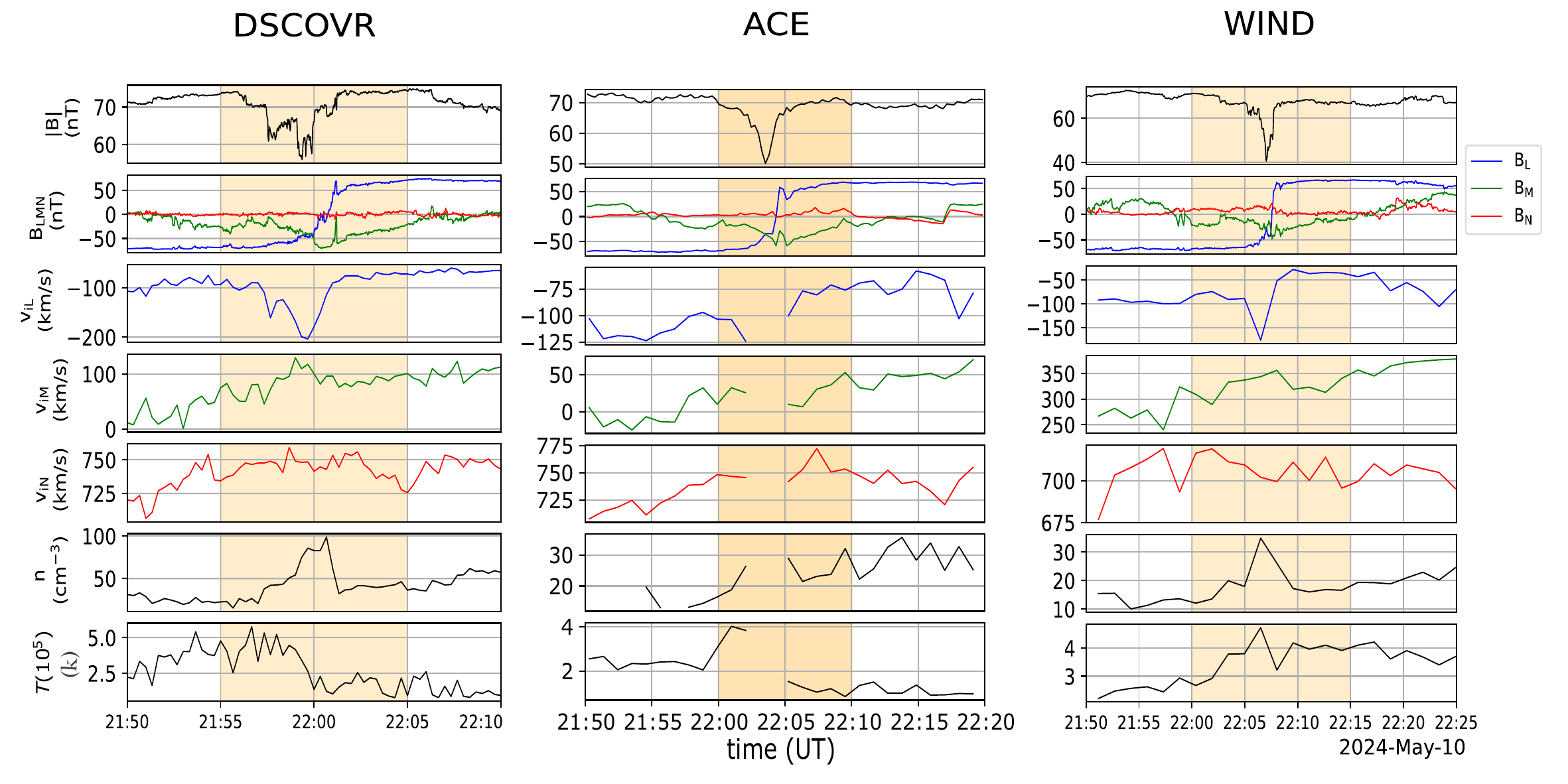}
\caption{Comparison of DSCOVR, ACE, and Wind data during the ICME flux rope interval. Panels display the IMF magnitude, its components, solar wind velocity components, density, and temperature in a boundary normal coordinate system (LMN). The gaps in ACE observations are due to the unavailability of observations.}
\label{figure:Reconnection}
\end{figure*}

Besides NASA's observatories at the Sun-Earth L1 point, the ADITYA-L1 mission of ISRO is the latest edition \citep{Tripathi_Chakrabarty_Nandi_Raghvendra_Prasad_Ramaprakash_Shaji_Sankarasubramanian_Satheesh_Thampi_Yadav_2022}. All spacecraft monitor the upstream solar wind conditions with high-precision instruments. Leveraging spacecraft positions along the GSE-y direction (dawn-dusk), we investigated the region of interest in figure~\ref{figure:summary} spatially. We comprehensively analyzed plasma data from DSCOVR, ACE, and Wind. Due to the unavailability of plasma data, Aditya L1 observations could not be used. Figure~\ref{figure:Reconnection} represents magnetic field and plasma moments data transformed into a boundary-normal frame using Minimum Variance Analysis (MVA\citep{Sonnerup_1967}. From 21:50 UT to 22:25 UT, all three spacecraft exhibited consistent observations with timing offsets due to separation of the spacecrafts along the GSEx direction (please refer to Appendix~\ref{sec:spec_pos}). A systematic decrease in the minimum magnetic field magnitude ($|B|_{min}$) and a sharp rotation in the $B_{L}$ component were observed along the dawn-dusk line. Concurrently, a localized dip followed by an increase in the L component of solar wind ion velocity ($V_{iL}$) was detected, with the width of the velocity dip decreasing progressively from DSCOVR to Wind. Further, density (n) and temperature (T) showed a localized order-of-magnitude increase across all three spacecraft. These observations indicate active magnetic reconnection and the presence of a current sheet (CS). The $V_{iL}$ enhancement near the magnetic reversal matches the reconnection outflow signature, with bulk flow speed reaching up to 200 km/s along L, twice the background velocity. Observations suggest that all three spacecraft have traversed different spatial regions within the same reconnection outflow. 

To quantify the reconnection outflow spatial scales and construct its geometry, plasma parameters were derived from in situ electron and proton plasma moments from DSCOVR, Wind, and ACE. The local plasma Debye length $\lambda_{d}$ was $\sim5$ meters ($\lambda_{d} = \sqrt{\epsilon K_{B} T_{e} / n_{e} e^2}$, where\, $K_{B},\,\epsilon,\, T_{e},\,n_{e}$,\, represents Boltzmann constant, medium permittivity, electron temperature and density). The ion inertial length ($d_{i}$) was 15 km and electron inertial length ($d_{e}$) falls in the range of 595-600 meter ($d_{i(e)}=c/\omega_{pi(e)}$, where $\omega_{pi(e)}$ and c denote the ion(electron) plasma frequency and speed of light). The electron and ion inertial lengths, or skin depths, define the characteristic scales beyond which electrons and ions decouple from the magnetic field. Given the high solar wind velocity, spacecraft velocities were deemed negligible. Therefore, using timing analysis, the width of the outflow jets was determined to be 26, 30, and 13 $R_{E}$ (where 1 $R_{E}$= 6317 km) for DSCOVR, ACE, and Wind, respectively. Notably, the $B_{L}$ component reversal and the outflow jet exhibit consistent directional alignment along -ve L (or GSEy), indicating the current sheet orientation is closely aligned with the GSE-y axis. These results imply the reconnection diffusion region location was duskward of the Wind spacecraft. Using fluxgate magnetometer data from Aditya-L1 spacecraft \citep{Yadav2025}, located 100 $R_E$ dawnward, the current sheet's spatial extent was estimated at $\sim$203 $R_E$. Although Aditya-L1's plasma data is unavailable for this exercise, its unique location enabled us to determine the maximum observable length of the reconnection outflow based on its magnetometer data. Therefore, the spatial separation of all the L1 observatories and near-simultaneous observations enables us to construct the reconnection geometry, which is explained in the discussion section.

\begin{figure*}[htbp]
\centering
\includegraphics[width=0.99\linewidth]{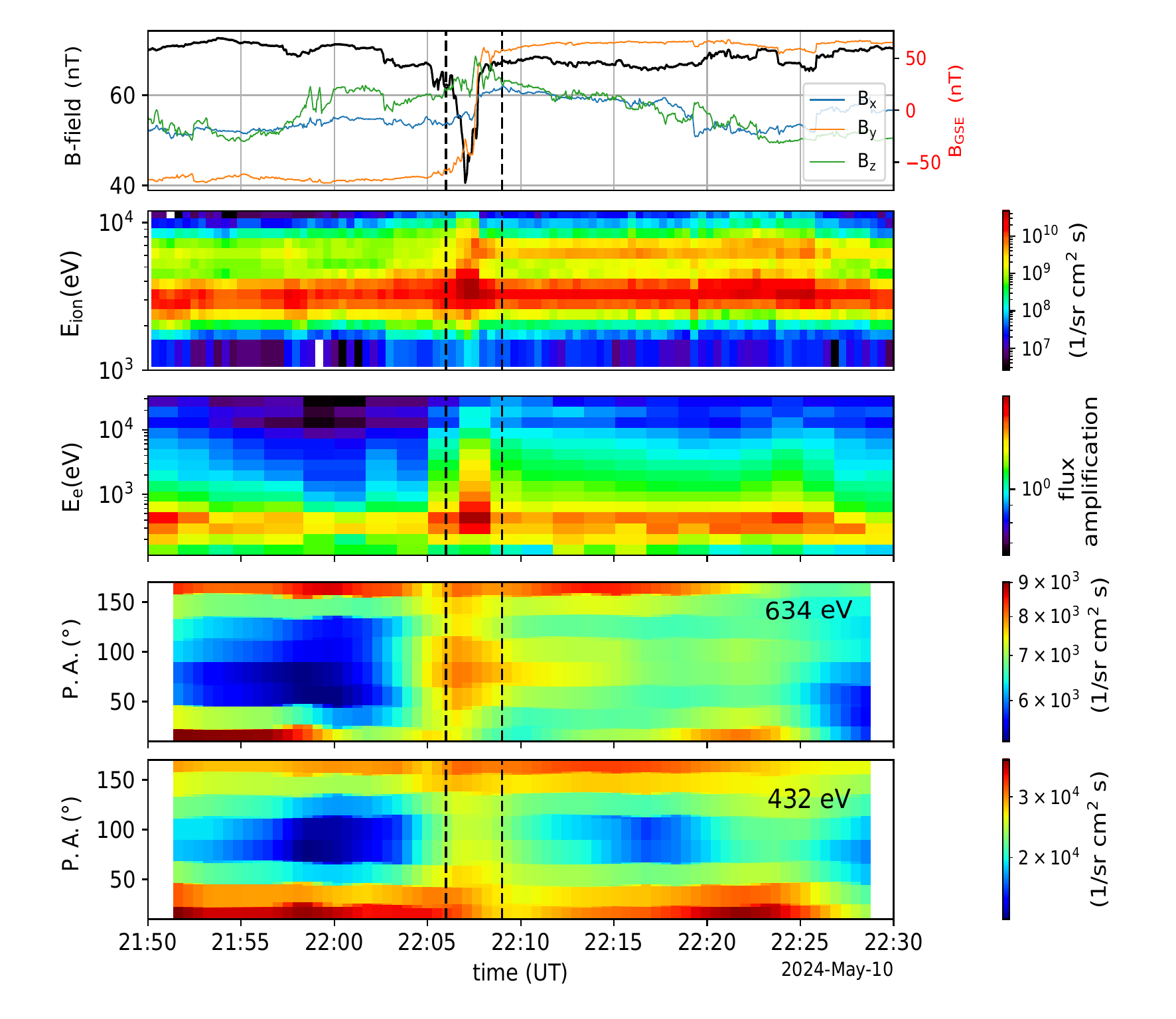}
\caption{This figure shows from the top timeseries data of: IMF magnitude (B-field) and components, solar wind ion energy spectrogram ($E_{ion}$), electron energy spectrogram ($E_{e}$, normalized by 21:00 UT flux, outside the MC), and solar wind suprathermal electron pitch angle distribution (634 \& 432 eV) respectively from  21:50 UT to 22:30 UT on May 10 from Wind spacecraft data. The other suprathermal electron channels ($>$ 60 eV) show similar behavior. The white bands represent data gaps. Two black vertical lines mark the region of interest.}
\label{figure:spectogram}
\end{figure*}

In magnetic reconnection, magnetic energy is efficiently converted into plasma kinetic energy, accelerating charged particles, particularly within the outflow jets, where particles can attain relativistic speeds. Using the Three-dimensional plasma and Energetic Particle Investigation Instrument \citep{lin1995three} aboard the Wind mission, the electron and ion omnidirectional energy flux spectrogram is displayed in figure~\ref{figure:spectogram}b. We observe a significant enhancement in ion energy, reaching up to 10 keV, compared to the background energy band of 2-4 keV. The electron energy spectrum also exhibits a noticeable flux amplification by an order of magnitude between 22:05 UT and 22:10 UT. Hence, it is evident that notable particle energization has taken place, coinciding with the $B_{min}$ position. Moreover, the ion flux displays a distinct band centered around 6-8 keV, which is twice the energy of observed protons and can be attributed to the presence of helium ions. Studies have shown that MC internal magnetic reconnections often disrupt the coherence of the flux tube \citep{farrugia2023magnetic}, leading to flux erosion \citep{pal2020flux}. \cite{carcaboso2020characterisation} have characterized the pitch-angle-distribution (PADs) of solar wind suprathermal electrons, highlighting their significance in diagnosing the magnetic topology. A key finding of the study depicts that bidirectional PADs indicate closed magnetic structures, while isotropic distributions suggest disconnection by magnetic reconnection or highly turbulent sheaths. In this present case, we observed high field-aligned electron flux throughout the region of interest, except at the $B_{min}$ point, which indicates isotropic PADs, suggesting a disconnected field line due to reconnection. These observations provide compelling evidence to conclude the presence of an ongoing magnetic reconnection within the flux rope and particle acceleration in its outflow regions.

\section{Discussion}\label{sec:Discussion}
ICME flux rope magnetic fields are often modeled as twisted circular cross-sections (see e.g.\cite{gomez2017sunward}); although it may not hold in reality \citep{al2011internal}. In 1974, J.B. Taylor showed that Reversed Field Pinches(RFP) relax to a minimum-energy Taylor state via magnetic reconnection \citep{taylor1974relaxation}, satisfying $\nabla \times \mathbf{B} = \lambda B$ with constant $\lambda$. If magnetic flux $\Phi = \int_A \vec{B} \cdot \vec{dn}$ and helicity $K = \int_\tau \vec{A}.\vec{B}\, d\tau$ (A is magnetic vector potential) are invariant in RFP geometry, relaxation occurs through reconnection \citep{taylor1986relaxation}. Approximately, solar flux ropes can be considered to originate in a force-free state \citep{burlaga1981magnetic,lepping1990magnetic}, though not always \citep{xie2023magnetic}. Deviations from the Taylor state in the interplanetary medium may trigger reconnection. Reconnection signatures inside ICME flux ropes have been observed previously by \cite{gosling2005direct,gosling2006petschek}. In a pioneering work, \cite{fermo2014magnetic} demonstrated the occurrence of reconnection in a stretched or elongated flux rope through numerical simulation, motivating our effort to link such evidence to the reconnection origin in this study. \\

\begin{figure*}[htbp]    
\centering
\includegraphics[width=0.99\linewidth]{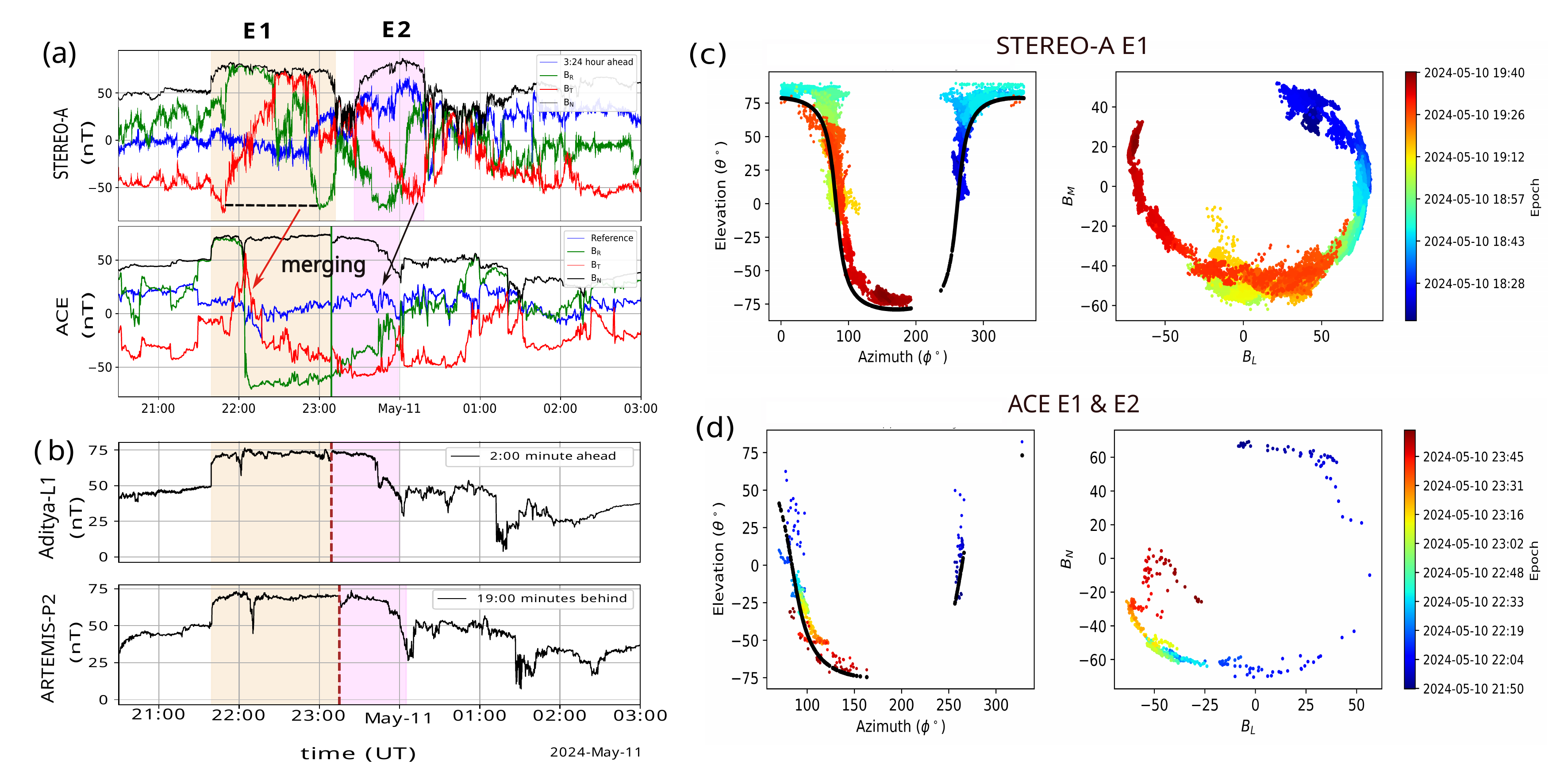}
\caption{The timeshifted magnetic field observation from STEREO-A to match the MC initiation at 21:39:22 UT with ACE in panel (a) in RTN coordinates, followed by ADITYA-L1 and THEMIS-C/ARTEMIS P2 observation of the reconnection, respectively. The areas marked in antique white and magenta are two successive Ejectas/MCs, E1 and E2. Panel (c) \& (d) describes the azimuth vs elevation angle and 2D magnetic field hodogram result from MVA analysis of STEREO-A and ACE magnetic field data corresponding to  E1 MC.}
\label{figure:timeshift}
\end{figure*}

The diverse spatial distribution of the spacecrafts (figure~\ref{figure:Spacecraft}) provided a unique opportunity for examining the spatiotemporal evolution of the MCs (MC identification criteria are presented in Appendix~\ref {sec:MC identification}). STEREO-A \citep{kaiser2007stereo}, positioned 1798 $R_{E}$ sunward ($GSE_{x}$) and 4907 $R_{E}$ dawnward (-$GSE_{y}$), observed the event 3.24 hour before ACE. Two distinct magnetic regions, E1 and E2 (antique white and magenta in figure~\ref{figure:timeshift}a), having similar IMF intensity, are separated by an interaction region (May 10, 19:40-20:06 UT) marked by abrupt magnetic fluctuations. 
Minimum Variance Analysis (MVA, details in Appendix~\ref{sec:MVA}) and 2D hodogram of STEREO-A data reveal that E1 exhibits a coherent, large-scale magnetic rotation, with a smooth color gradient in the $\phi-\theta$ plot (figure~\ref{figure:timeshift}c) and an elevation change from $+75^\circ$ to $-75^\circ$ across a full $ 0^\circ-360^\circ$ azimuth span. This suggests a well-developed flux rope with an almost perpendicular orientation ($\theta_{max}= 89.98^\circ$). The hodogram's arc pattern further supports this interpretation. In contrast, E2 displays a different orientation ($\theta_{max}= 63.68^\circ$) and reduced coherence, possibly indicating a distorted structure (not shown in plot). However, the lack of STEREO-A plasma data precludes definitive classification. Notably, E1 shows no significant sharp decrease in IMF intensity.\\ 

\begin{figure*}[htbp]
\centering
\includegraphics[width=0.9\textwidth]{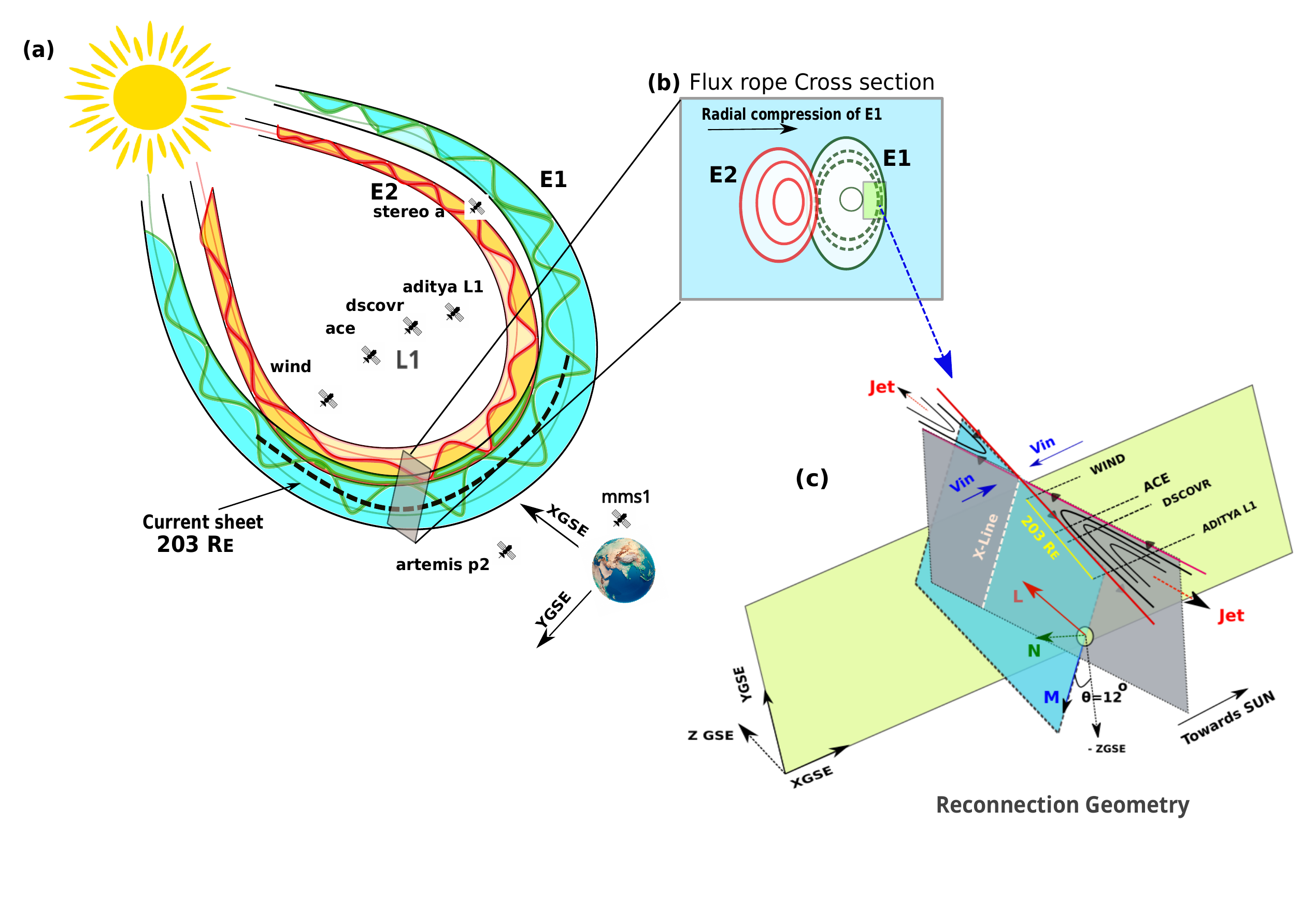}
\caption{Artistic illustration of the ICME-ICME interaction and embedded magnetic reconnection site. Panel (a) describes the interaction of two ICME MCs on May 10, 2024. Two prominent MCs (E1 \& E2) are marked with green and yellow, respectively, in the GSE coordinate system, with all the tentative positions of the spacecrafts (Appendix~\ref{sec:spec_pos}). The internal magnetic reconnection current sheet is marked in a black dashed line inside E1 MC spanning 203 $R_{E}$ along the dawn-dusk direction. Panel (b) represents the cross-sectional view of the reconnection site, indicating the structural deformation of E1 due to intense compression from the trailing MC E2. Panel (c) displays the 3D reconnection geometry in the GSE and local Boundary normal coordinate system (LMN, Appendix~\ref{sec:lmn}). All spacecraft around the L1 point probed the same reconnection outflow jet. The ecliptic plane is visualized in green, while the sky blue plane represents the plane containing the reconnection X-LINE, which is along the M direction, inclined to - ZGSE at an angle of $12^\circ$. This sketch is for illustration purposes only and not to scale.}
\label{figure:geometry}
\end{figure*}

At L1, ACE observed a similar $B_{N}$ and $B_{T}$ variation to STEREO-A, though in a more compressed form, with the two structures (E1 and E2) appearing substantially overlapped. In figure~\ref{figure:timeshift}a, the STEREO-A MC interval (black dotted horizontal line) corresponds to a compressed section in the ACE data (red arrow). At the same time, the trailing portion of E2 remains detectable visually (black arrow). Comparable durations of E1 and E2 in both datasets suggest interacting or sequentially merging flux ropes. MVA and 2D hodogram analysis of ACE data (21:40-23:59 UT, May 10; figure~\ref {figure:timeshift}d) reveal a discontinuous $\phi-\theta$ relation, contrasting with the smooth STEREO-A flux rope signature. The limited elevation range, partial azimuth coverage, and temporal discontinuities may imply intermittent reconnection, though alternative explanations include a partially developed flux rope or short-duration crossing. The pronounced $|\vec{B}|$ drop supports the presence of an internal reconnection region. Consistent findings are obtained from other L1 spacecraft (see Appendix \ref{sec:MVA}). Between L1 and ARTEMIS-P2, $\theta_{max}$ further decreases from $\sim80^\circ$ to $\sim54.79^\circ$, while $\lambda_{2}/\lambda_{3}$ increases from 16 to 40 (table \ref{tab:time_intervals}), indicating a rotation of the boundary normal relative to the upstream field and a more well-defined planar boundary. This behavior is consistent with, though not uniquely diagnostic of, progressive flattening (\lq{}pancaking\rq{}) of the magnetic cloud during propagation\citep{raghav2023possible}.\\ 

The cross-section of ICME MC rapidly flattens or pancakes, losing its coherence, owing to the radial expansion \citep{owens2017coronal,owens2005characteristic}. Moreover, large-scale current sheets and subsequent magnetic reconnection can trigger inside MCs due to the deformation \citep{owens2009formation,fermo2014magnetic}. Therefore, we argue that due to the interaction and merging of E1 \& E2 between STEREO-A and L1 position, E1 became compressed, which might have led to deviation from the Taylor state and induced internal reconnection within E1, specifically involving the $B_{T}$ component of its internal surface magnetic field, facilitated by the formation of a current sheet. This internal MC reconnection was observed by Aditya L1, two minutes before ACE, and subsequently by DSCOVR, ACE, Wind, ARTEMIS P2 (figure~\ref{figure:timeshift}b). This temporal sequence indicates the quasi-steady nature of the ongoing reconnection for at least 35 minutes, including MMS observation near the magnetopause. An artistic illustration describing this observational scenario is presented in figure~\ref {figure:geometry}a, where the spacecraft positions used in the analysis are schematically represented in figure~\ref{figure:geometry}a. Figure~\ref{figure:geometry}b provides a cross-sectional view of the intricate flux rope interaction and the ongoing magnetic reconnection between the surface magnetic fields. The reconnection geometry is shown in GSE coordinates with all the L1 spacecraft crossings schematically in figure~\ref {figure:geometry}c. \\

While the above physical picture explains the observations, the solution is not unique and depends on underlying assumptions. A possible alternative is a geometric effect: STEREO-A and ACE were separated in the L($GSE_{y}$) direction by over 4000 $R_{E}$. Given the propagation of the two MCs, merging or interaction may have occurred at the Earth-directed section, distant from STEREO-A. The absence of STEREO-A plasma data limits detailed analysis of E1 and E2. If E1 and E2 at ACE represent a single flux rope, the interaction or merging hypothesis would be invalid, favouring deformation and compression instead. However, the Wind spacecraft electron and ion omnidirectional energy, along with suprathermal electron fluxes, reveal contrasting evidence between E1 AND E2 (figure~\ref{figure:Wind_fluxrope}), indicating disparate plasma behaviour and supporting the merging of two MCs. Although magnetic field signatures alone cannot separate the structures, plasma data provide a complementary perspective. Moreover, a solar wind bulk speed increase behind the E1-E2 boundary (figure~\ref{figure:summary}Ie) suggests compression from ICME-ICME interaction, inhibiting E1's expansion \citep{russell2005defining}. Thus, multispacecraft observations indicate reconnection initiated by merging MCs via current sheet formation, and its impact caused rapid changes in the magnetosphere-ionosphere system. Future numerical modeling could further substantiate this interpretation.   

\section{Acknowledgement}
\begin{acknowledgments}
The authors express their sincere gratitude to the ACE, DSCOVR, Wind, STEREO, MMS, and THEMIS mission teams of NASA and ISRO's Aditya-L1 for their dedicated efforts in operating the missions and for providing high-quality scientific data in the public domain. S.B. appreciates the financial support provided by the Indian Space Research Organization through a research grant. The magnetic field and plasma data of NASA's ACE, Wind, DSCOVR, and ARTEMIS-P2 \& STEREO-A  are publicly available on the following website: \url{https://cdaweb.gsfc.nasa.gov/}. One can visit the science gateways of all the spacecraft mentioned above to check the current data availability. The ADITYA L1 mission data are available at: \url{https://www.issdc.gov.in/adityal1.html}. The Python-based Space Physics Environment Data Analysis Software (pySPEDAS) (\url{https://github.com/spedas/pyspedas}) has been used to analyze and plot the mission data. 
\end{acknowledgments}

\appendix

\section{Methods}

\subsection{Boundary Normal Coordinate (LMN) determination.}
\label{sec:lmn}
Minimum Variance Analysis (\cite{Sonnerup_1967}) has been used to transform the in-situ data from the GSE coordinate system to the boundary normal coordinate system. In the GSE coordinate basis, the new coordinate system components are L = [0.61615, 0.98312, 0.08625], M = [-0.1717, 0.11404, -0.97851], N = [-0.97183, 0.14306, 0.18727]. The pySPEDAS module has been used to determine the minimum variance matrix and then applied to the data to transform them from GSE to LMN coordinates. For further details, please refer to the GitHub tutorials of pySPEDAS (link provided in Acknowledgement)\\
\begin{table}[h!]
\centering
\begin{tabular}{lccc}
\toprule
\textbf{LMN} & \textbf{Angle to XGSE} & \textbf{Angle to YGSE} & \textbf{Angle to ZGSE} \\ 
\midrule
L    &  $80.71^{\circ}$ & $10.54^{\circ}$ & $85.05^{\circ}$ \\ \hline
M    & $99.89^{\circ}$ & $83.45^{\circ}$ & $168.36^{\circ}$ \\ \hline
N    & $166.36^{\circ}$ & $81.77^{\circ}$ & $79.20^{\circ}$ \\ 
\bottomrule
\end{tabular}
\caption{Orientation of Reconnection LMN to GSE coordinate}
\label{tab:reconnectionLMN}
\end{table}

\subsection{Minimum Varience analysis and 2D hodogram.}
\label{sec:MVA}
When magnetic field $\vec{B} = (B_{x},\, B_{y},\, B_{z}) = (B cos\theta\,cos\phi,\, Bcos\theta\,sin\phi,\, Bsin\theta)$, are parallel to a plane having normal $n = (n_{x},\,n_{y},\,n_{z})$ (where $\theta\, and\,\phi$ are azimuthal and inclination angles of magnetic field vector respectively), the relation between $\theta$ and $\phi$ is given by \cite{nakagawa1989planar,nakagawa1993solar,palmerio2016planar} \\
\begin{equation}
   n_{x}cos\theta cos\phi\,+\,n_{y}cos\theta sin\phi \,+\, n_{z}sin\theta = 0 
\end{equation}
The above curve fitting to the measured $\theta$ and $\phi$ distribution in $\phi-\theta$ space indicates the presence of a Planar Magnetic Structure (PMS). To check the planarity of the vector in 2D, $|B_{n}|/B$ is calculated, where B is the magnitude of the IMF and $B_{n}$ is a component of the magnetic field normal to the PMS plane. The PMS will be in a perfect plane when $B_{n}\approx 0$. Therefore, a low value of $|B_{n}|/B$ indicates the vector is parallel to a plane.\\
The Minimum Variance Analysis (MVA) technique is used to identify the orientation and the normal of a magnetic field structure. The MVA analysis for the selected region estimates three new directions as maximum ($B_{L}$), intermediate ($B_{M}$), and minimum ($B_{N}$) variances corresponding to maximum ($\lambda_{1}$), intermediate ($\lambda_{2}$), and minimum ($\lambda_{3}$) eigenvalues and three eigenvectors ($\hat{n_{1}},\,\hat{n_{2}},\, \hat{n_{3}}$), respectively \citep{Sonnerup_1967, palmerio2016planar}. Here, $\hat{n_{3}} = \hat{n}$ represents the normal to the plane. The ratio of the intermediate to minimum eigenvalue ($R = \lambda_{2}/\lambda_{1}$) indicates the efficiency of the MVA technique \citep{shaikh2018identification}. The MVA time intervals are shown below. The identification of planar magnetic structures is presented in detail in \cite{nakagawa1989planar,neugebauer1993origins,palmerio2016planar,shaikh2018identification}. Generally, smooth rotation of magnetic field components i.e arc or semicircular pattern in one of the planes ($B_{L}$ - $B_{M}$, $B_{L}$ - $B_{N}$, $B_{M}$ - $B_{N}$) indicates rotational structures of magnetic fields \citep{khabarova2015small,shaikh2017presence} in 2D hodogram plot.

\section{Magnetic Cloud selection criteria and intervals}
\label{sec:MC identification}
In this study, we utilized the basic criteria for Magnetic cloud identification using magnetic field and plasma parameters. 
The STEREO-A MC E1 and E2 have been identified based on (1) enhanced magnetic field compared to surroundings ($>10$ nT ); (2) smooth rotation of magnetic field components over a large angle. Further, E1 and E2 were well separated by a highly fluctuating magnetic field, indicating an interaction region. However, due to the non-availability of plasma data from STEREO-A, we are unable to check plasma beta and temperature, which are essential parameters for MC identification. Still, Minimum Variance Analysis and 2D hodogram analysis enabled us to identify the E1 \& E2 (with $\theta_{max}$=89.98 and 63.68, respectively) region to be a well-developed and deformed flux rope, respectively.\\
At L1,  ACE observation indicates a single MC instead of E1 and E2 separately. With the help of the magnetic field of ACE only, we cannot distinguish them. But there is specific observational evidence that highlights the merging of E1 and E2 at L1. These are:\\
\textbf{1. Using magnetic field data and comparing the time duration of E1 and E2 between STEREO-A and ACE}:  quantitatively, the time duration of STEREO-A E1 \& E2 and their ACE counterpart is similar (approximately  2:20 hours). This also suggests the expansion of E1 due to high compression. Moreover, the magnetic field orientation (2024-05-10/21:39:00 to 23:59:00 UT) resembles that of STEREO-A E1 in a compressed manner. The trailing portion of E2 can also be identifiable (figure 4).\\ 
\textbf{2. From Plasma Observation using Wind}: The above explanation is not sufficient to identify the  E1 and E2 regions at L1. To further support our inference, we utilized electron, ion energy flux, and suprathermal electron pitch angle distributions(PADs) data. We notice distinct changes in the spectrograms and PADs after 2024-05-10/23:09:00 UT (boundary between E1 and E2).\\ 
All other observatories at L1 show similar evidence, and MC boundaries have been identified based on timing analysis. Therefore, we concluded the merging of two ICME MCs at L1 and used the E1 \& E2 intervals for MVA analysis (apart from STEREO-A).

\begin{table}[h!]
\centering
\begin{tabular}{|c|c|c|c|}
\toprule

\textbf{Start (UT)} & \textbf{End (UT)} & \textbf{Event} & \textbf{MVA analysis results} \\ \hline
\midrule
2024-05-10/18:15:00 & 2024-05-10/19:40:00    & STEREO-A MC E1 & $\theta_{max} = 89.98$, $<B_{n}>/<B> = 0.41$, $\lambda_{2}/\lambda_{3} = 63.31$ \\ \hline
2024-05-10/20:06:00 & 2024-05-10/20:53:00   & STEREO-A MC E2 & $\theta_{max} = 63.68$, $<B_{n}>/<B> = 0.52$, $\lambda_{2}/\lambda_{3} = 3.08$\\ \hline
2024-05-10/21:37:00 & 2024-05-10/23:57:00   & Aditya-L1 MC E1 \& E2 & $\theta_{max} = 73.34$, $<B_{n}>/<B> = 0.76$, $\lambda_{2}/\lambda_{3} = 37.71$\\ \hline
2024-05-10/21:36:00 & 2024-05-10/23:55:35    & DSCOVR  MC E1 \& E2 & $\theta_{max} = 80.11$, $<B_{n}>/<B> = 0.66$, $\lambda_{2}/\lambda_{3} = 31.02$ \\ \hline
2024-05-10/21:39:00 & 2024-05-10/23:59:00    & ACE MC E1 \& E2  & $\theta_{max} = 82.24$, $<B_{n}>/<B> = 0.68$, $\lambda_{2}/\lambda_{3} = 26.96$\\ \hline
2024-05-10/21:45:00 & 2024-05-10/23:40:00   & Wind MC E1 \& E2  & $\theta_{max} = 82.98$, $<B_{n}>/<B> = 0.70$, $\lambda_{2}/\lambda_{3} = 16.53$\\ \hline
2024-05-10/21:58:00 & 2024-05-11/00:19:00   & Artemis-P2 MC E1\& E2 & $\theta_{max} = 54.79$, $<B_{n}>/<B> = 0.69$, $\lambda_{2}/\lambda_{3} = 40.36$ \\ 
\bottomrule
\end{tabular}
\caption{Time intervals of MVA analysis. The columns represent the start and end time of E1 and E2 MCs detected by the spacecrafts mentioned, and the Minimum variance analysis results, respectively. $\theta_{max}$,  $<B_{n}>/<B>$, and $\lambda_{2}/\lambda_{3}$ provide a rough orientation of the flux rope, presence of magnetic field along the direction normal to the flux rope boundary, and the quality or reliability of the MVA result, respectively. }
\label{tab:time_intervals}
\end{table}

\section{Wind Magnetic Cloud flux and Pitch Angle}
\begin{figure*}[htbp]
\centering
\includegraphics[width=0.9\textwidth]{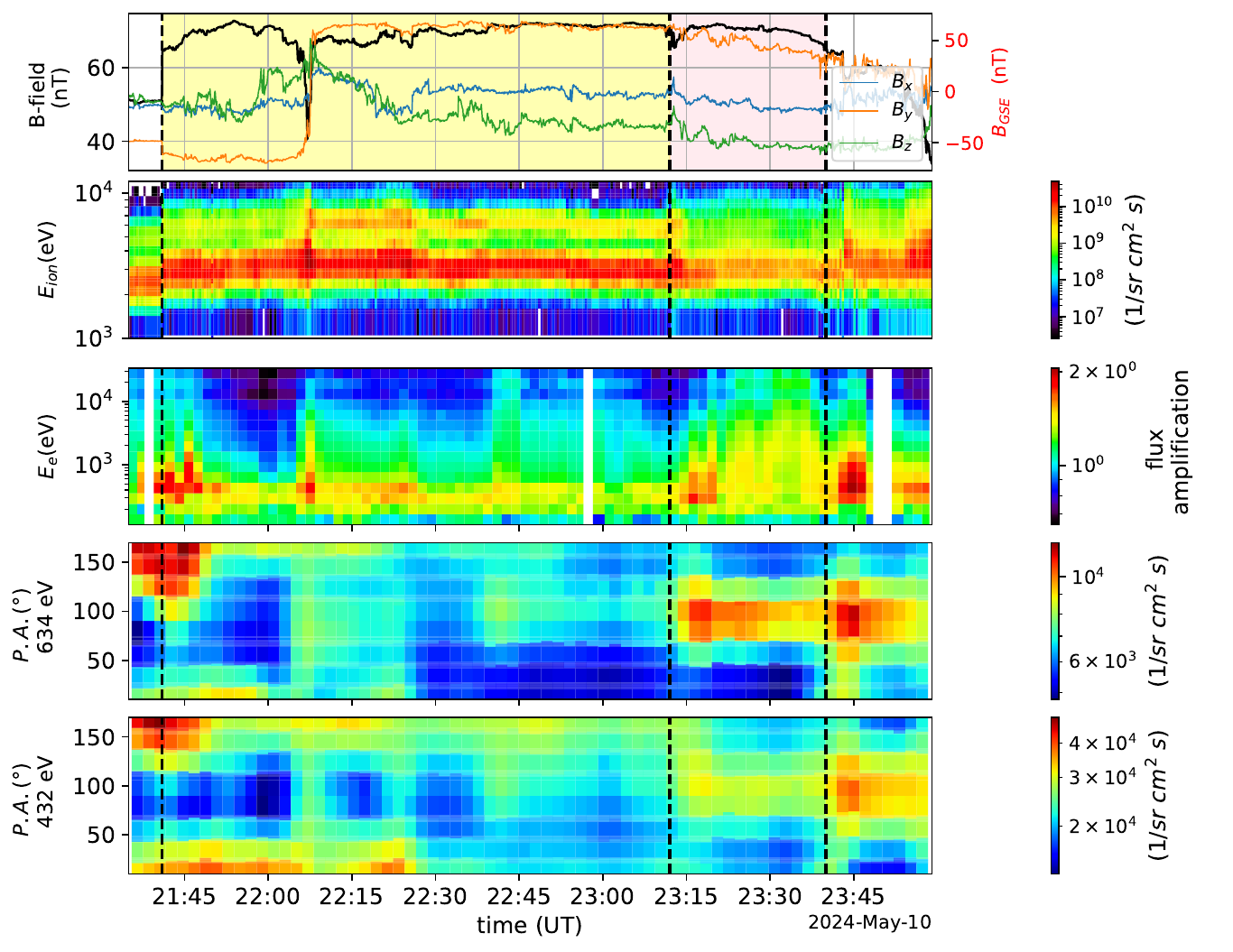}
\caption{Wind spacecraft observation of the complex magnetic cloud structure. Panels display the magnetic field, ion, and electron omnidirectional energy flux, as well as the suprathermal electron pitch angle distribution, in two separate channels. The yellow and pink shaded regions represent two different flux ropes. }
\label{figure:Wind_fluxrope}
\end{figure*}

\section{\textbf{Spacecraft Positions}}
\label{sec:spec_pos}
\begin{figure*}[htbp]
\centering
\includegraphics[width=0.9\textwidth]{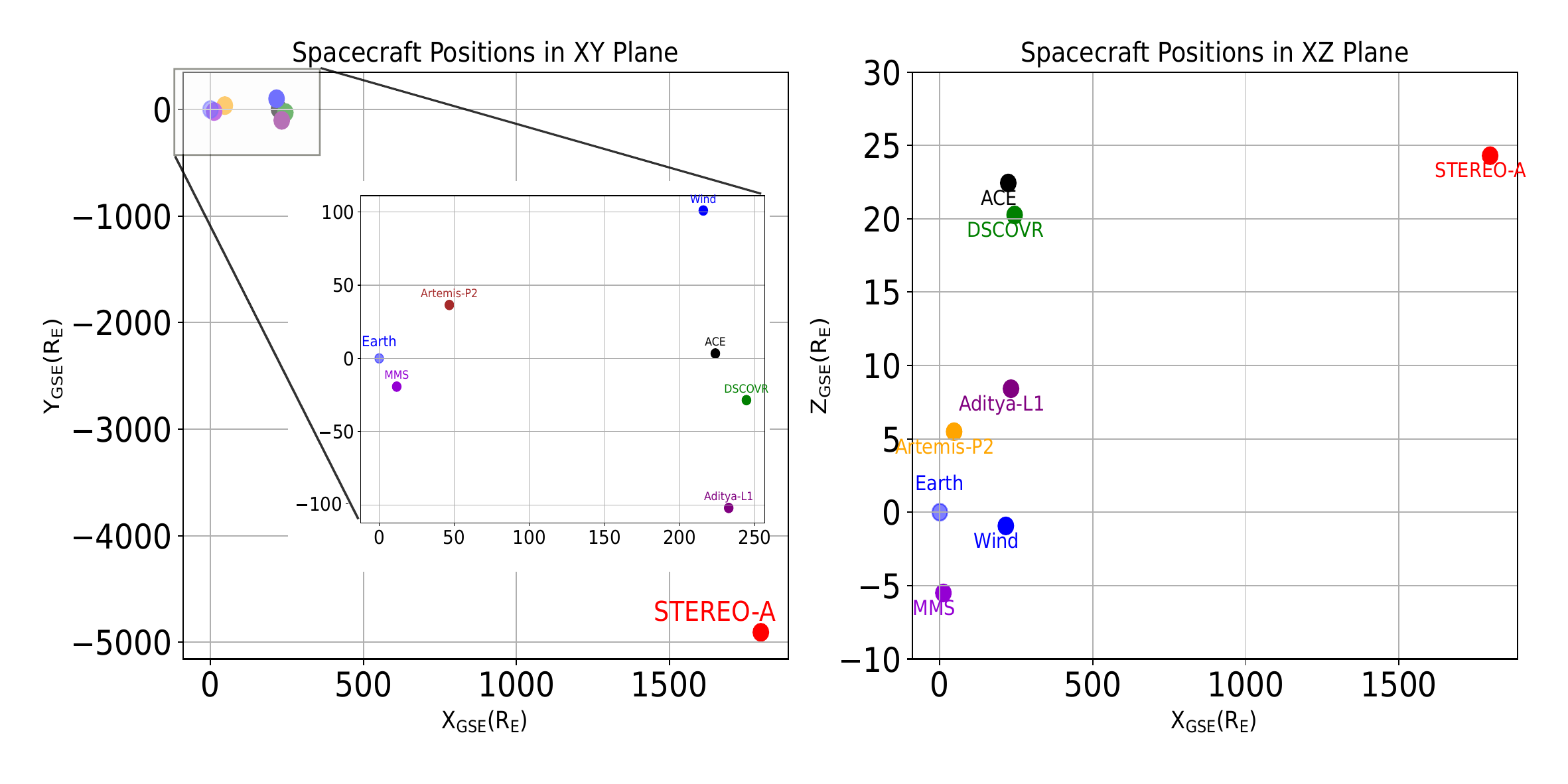}
\caption{Spacecraft positions in Geocentric Solar Ecliptic (GSE) Coordinate System on 10th May 2025 at 22:00 UT}
\label{figure:Spacecraft}
\end{figure*}

\begin{table}[h!]
\centering
\begin{tabular}{lccc}
\toprule
\textbf{Spacecraft} & \textbf{XGSE ($R_{E}$)} & \textbf{YGSE ($R_{E}$)} & \textbf{ZGSE ($R_{E}$)} \\
\midrule
ACE     & 223.77 & 3.35 & 22.44 \\ \hline
Wind    & 215.75 & 100.99 & -0.93 \\ \hline
DSCOVR  & 244.46 & -28.49 & 20.26 \\ \hline
ARTEMIS P2 & 46.69 & 36.48 & 5.49 \\ \hline
Aditya-L1 & 232.67 & -102.12 & 8.42 \\ \hline
STEREO-A & 1798.6 & - 4906.5 & 24.3 \\ \hline
MMS  & 11.66 & -19.27 & -5.50 \\
\bottomrule
\end{tabular}
\caption{GSE coordinates of spacecrafts on 10th May 2025 at 22:00 UT}
\label{tab:spacecraft_coords}
\end{table}

\bibliography{bibliography}

\begin{thebibliography}{}
\expandafter\ifx\csname natexlab\endcsname\relax\def\natexlab#1{#1}\fi
\providecommand{\url}[1]{\href{#1}{#1}}
\providecommand{\dodoi}[1]{doi:~\href{http://doi.org/#1}{\nolinkurl{#1}}}
\providecommand{\doeprint}[1]{\href{http://ascl.net/#1}{\nolinkurl{http://ascl.net/#1}}}
\providecommand{\doarXiv}[1]{\href{https://arxiv.org/abs/#1}{\nolinkurl{https://arxiv.org/abs/#1}}}

\bibitem[{Acu{\~n}a {et~al.}(2008)Acu{\~n}a, Curtis, Scheifele, Russell, Schroeder, Szabo, \& Luhmann}]{acuna2008stereo}
Acu{\~n}a, M., Curtis, D., Scheifele, J., {et~al.} 2008, Space Science Reviews, 136, 203, \dodoi{10.1007/978-0-387-09649-0_8}

\bibitem[{Al-Haddad {et~al.}(2011)Al-Haddad, Roussev, M{\"o}stl, Jacobs, Lugaz, Poedts, \& Farrugia}]{al2011internal}
Al-Haddad, N., Roussev, I.~I., M{\"o}stl, C., {et~al.} 2011, The Astrophysical Journal Letters, 738, L18, \dodoi{10.1088/2041-8205/738/2/L18}

\bibitem[{Angelopoulos(2008)}]{angelopoulos2008themis}
Angelopoulos, V. 2008, Space Science Reviews, 141, 5, \dodoi{10.1007/978-0-387-89820-9_2}

\bibitem[{Ashna {et~al.}(2024)Ashna, Bhaskar, Manju, \& Sini}]{ashna2024solar}
Ashna, V., Bhaskar, A., Manju, G., \& Sini, R. 2024, Journal of Geophysical Research: Space Physics, 129, e2023JA031687, \dodoi{10.1029/2023ja031687}

\bibitem[{Auster {et~al.}(2008)Auster, Glassmeier, Magnes, Aydogar, Baumjohann, Constantinescu, Fischer, Fornacon, Georgescu, Harvey, {et~al.}}]{auster2008themis}
Auster, H., Glassmeier, K., Magnes, W., {et~al.} 2008, Space science reviews, 141, 235, \dodoi{10.1007/978-0-387-89820-9_11}

\bibitem[{Burlaga {et~al.}(1981)Burlaga, Sittler, Mariani, \& Schwenn}]{burlaga1981magnetic}
Burlaga, L., Sittler, E., Mariani, F., \& Schwenn, a.~R. 1981, Journal of Geophysical Research: Space Physics, 86, 6673, \dodoi{10.1029/ja086ia08p06673}

\bibitem[{Burt \& Smith(2012)}]{burt2012deep}
Burt, J., \& Smith, B. 2012, IEEE, 1

\bibitem[{Carcaboso {et~al.}(2020)Carcaboso, G{\'o}mez-Herrero, Lara, Hidalgo, Cernuda, \& Rodr{\'\i}guez-Pacheco}]{carcaboso2020characterisation}
Carcaboso, F., G{\'o}mez-Herrero, R., Lara, F.~E., {et~al.} 2020, Astronomy \& Astrophysics, 635, A79, \dodoi{10.1051/0004-6361/201936601}

\bibitem[{Chakrabarty {et~al.}(2017)Chakrabarty, Hui, Rout, Sekar, Bhattacharyya, Reeves, \& Ruohoniemi}]{chakrabarty2017role}
Chakrabarty, D., Hui, D., Rout, D., {et~al.} 2017, Journal of Geophysical Research: Space Physics, 122, 2574, \dodoi{10.1002/2016ja022781}

\bibitem[{Farrugia {et~al.}(2023)Farrugia, Vasquez, Lugaz, Al-Haddad, Richardson, Davies, Winslow, Zhuang, Scolini, Torbert, {et~al.}}]{farrugia2023magnetic}
Farrugia, C., Vasquez, B., Lugaz, N., {et~al.} 2023, The Astrophysical Journal, 953, 15, \dodoi{10.3847/1538-4357/acdcf7}

\bibitem[{Fermo {et~al.}(2014)Fermo, Opher, \& Drake}]{fermo2014magnetic}
Fermo, R.~L., Opher, M., \& Drake, J.~F. 2014, Physical Review Letters, 113, 031101, \dodoi{10.1103/physrevlett.113.031101}

\bibitem[{Fu {et~al.}(2025)Fu, Fu, Zhang, Yu, \& Cao}]{fu2025compression}
Fu, W., Fu, H., Zhang, W., Yu, Y., \& Cao, J. 2025, Geophysical Research Letters, 52, e2024GL114040, \dodoi{10.1029/2024gl114040}

\bibitem[{G{\'o}mez-Herrero {et~al.}(2017)G{\'o}mez-Herrero, Dresing, Klassen, Heber, Temmer, Veronig, Bu{\v{c}}{\'\i}k, Hidalgo, Carcaboso, Blanco, {et~al.}}]{gomez2017sunward}
G{\'o}mez-Herrero, R., Dresing, N., Klassen, A., {et~al.} 2017, The Astrophysical Journal, 840, 85, \dodoi{10.3847/1538-4357/aa6c5c}

\bibitem[{Gonzalez {et~al.}(1994)Gonzalez, Joselyn, Kamide, Kroehl, Rostoker, Tsurutani, \& Vasyliunas}]{gonzalez1994geomagnetic}
Gonzalez, W., Joselyn, J.-A., Kamide, Y., {et~al.} 1994, Journal of Geophysical Research: Space Physics, 99, 5771, \dodoi{10.1029/93ja02867}

\bibitem[{Gonzalez \& Parker(2016)}]{gonzalez2016magnetic}
Gonzalez, W., \& Parker, E. 2016, Astrophysics and space science library, 427, 10, \dodoi{10.1007/978-3-319-26432-5}

\bibitem[{Gonzalez {et~al.}(2011)Gonzalez, Echer, Tsurutani, Cl{\'u}a~de Gonzalez, \& Dal~Lago}]{gonzalez2011interplanetary}
Gonzalez, W.~D., Echer, E., Tsurutani, B.~T., Cl{\'u}a~de Gonzalez, A.~L., \& Dal~Lago, A. 2011, Space science reviews, 158, 69, \dodoi{10.1007/s11214-010-9715-2}

\bibitem[{Gopalswamy(2016)}]{gopalswamy2016history}
Gopalswamy, N. 2016, Geoscience Letters, 3, 1, \dodoi{10.1186/s40562-016-0039-2}

\bibitem[{Gosling {et~al.}(2006)Gosling, Eriksson, Skoug, McComas, \& Forsyth}]{gosling2006petschek}
Gosling, J., Eriksson, S., Skoug, R.~M., McComas, D., \& Forsyth, R. 2006, The Astrophysical Journal, 644, 613, \dodoi{10.1086/503544}

\bibitem[{Gosling {et~al.}(1974)Gosling, Hildner, MacQueen, Munro, Poland, \& Ross}]{gosling1974mass}
Gosling, J., Hildner, E., MacQueen, R., {et~al.} 1974, Journal of Geophysical Research, 79, 4581, \dodoi{10.1029/ja079i031p04581}

\bibitem[{Gosling {et~al.}(1973)Gosling, Pizzo, \& Bame}]{Gosling1973Anomalously}
Gosling, J., Pizzo, V., \& Bame, S.~J. 1973, Journal of Geophysical Research, 78, 2001, \dodoi{10.1029/ja078i013p02001}

\bibitem[{Gosling {et~al.}(2005)Gosling, Skoug, McComas, \& Smith}]{gosling2005direct}
Gosling, J., Skoug, R.~M., McComas, D., \& Smith, C. 2005, Journal of Geophysical Research: Space Physics, 110, \dodoi{10.1029/2004ja010809}

\bibitem[{Hapgood(2017)}]{hapgood2017space}
Hapgood, M. 2017, Space weather (IOP Publishing), \dodoi{10.1088/978-0-7503-1372-8ch1}

\bibitem[{Hayakawa {et~al.}(2020)Hayakawa, Ribeiro, Vaquero, Gallego, Knipp, Mekhaldi, Bhaskar, Oliveira, Notsu, Carrasco, Caccavari, Veenadhari, Mukherjee, \& Ebihara}]{Hayakawa_2020}
Hayakawa, H., Ribeiro, P., Vaquero, J.~M., {et~al.} 2020, The Astrophysical Journal Letters, 897, L10, \dodoi{10.3847/2041-8213/ab6a18}

\bibitem[{Hayakawa {et~al.}(2025)Hayakawa, Ebihara, Mishev, Koldobskiy, Kusano, Bechet, Yashiro, Iwai, Shinbori, Mursula, {et~al.}}]{hayakawa2025solar}
Hayakawa, H., Ebihara, Y., Mishev, A., {et~al.} 2025, The Astrophysical Journal, 979, 49, \dodoi{10.3847/1538-4357/ad9335}

\bibitem[{Hirshberg {et~al.}(1970)Hirshberg, Alksne, Colburn, Bame, \& Hundhausen}]{Hirshberg1970Observation}
Hirshberg, J., Alksne, A., Colburn, D., Bame, S., \& Hundhausen, A. 1970, Journal of Geophysical Research, 75, 1, \dodoi{10.1029/ja075i001p00001}

\bibitem[{Holappa {et~al.}(2020)Holappa, Asikainen, \& Mursula}]{holappa2020explicit}
Holappa, L., Asikainen, T., \& Mursula, K. 2020, Geophysical Research Letters, 47, e2019GL086676, \dodoi{10.1029/2019gl086676}

\bibitem[{Holappa {et~al.}(2021)Holappa, Robinson, Pulkkinen, Asikainen, \& Mursula}]{holappa2021explicit}
Holappa, L., Robinson, R., Pulkkinen, A., Asikainen, T., \& Mursula, K. 2021, Journal of Geophysical Research: Space Physics, 126, e2021JA029202, \dodoi{10.1029/2021ja029202}

\bibitem[{Kaiser \& Adams(2007)}]{kaiser2007stereo}
Kaiser, M.~L., \& Adams, W.~J. 2007, IEEE, 1, \dodoi{10.1109/aero.2007.352745}

\bibitem[{Kaiser {et~al.}(2008)Kaiser, Kucera, Davila, St.~Cyr, Guhathakurta, \& Christian}]{kaiser2008stereo}
Kaiser, M.~L., Kucera, T., Davila, J., {et~al.} 2008, Space Science Reviews, 136, 5, \dodoi{10.1007/978-0-387-09649-0_2}

\bibitem[{Khabarova {et~al.}(2015)Khabarova, Zank, Li, Le~Roux, Webb, Dosch, \& Malandraki}]{khabarova2015small}
Khabarova, O., Zank, G., Li, G., {et~al.} 2015, The Astrophysical Journal, 808, 181, \dodoi{10.1088/0004-637x/808/2/181}

\bibitem[{Khabarova {et~al.}(2021)Khabarova, Malandraki, Malova, Kislov, Greco, Bruno, Pezzi, Servidio, Li, Matthaeus, {et~al.}}]{khabarova2021current}
Khabarova, O., Malandraki, O., Malova, H., {et~al.} 2021, Space Science Reviews, 217, 38, \dodoi{10.1007/s11214-021-00814-x}

\bibitem[{Khabarova {et~al.}(2016)Khabarova, Zank, Li, Malandraki, le~Roux, \& Webb}]{khabarova2016small}
Khabarova, O.~V., Zank, G.~P., Li, G., {et~al.} 2016, The Astrophysical Journal, 827, 122, \dodoi{10.3847/0004-637x/827/2/122}

\bibitem[{Kilpua {et~al.}(2017)Kilpua, Koskinen, \& Pulkkinen}]{kilpua2017coronal}
Kilpua, E., Koskinen, H.~E., \& Pulkkinen, T.~I. 2017, Living Reviews in Solar Physics, 14, 1, \dodoi{10.1007/s41116-017-0009-6}

\bibitem[{Klein \& Burlaga(1982)}]{klein1982interplanetary}
Klein, L., \& Burlaga, L. 1982, Journal of Geophysical Research: Space Physics, 87, 613, \dodoi{10.1029/ja087ia02p00613}

\bibitem[{Kruparova {et~al.}(2024)Kruparova, Krupar, Szabo, Lario, Nieves-Chinchilla, \& Oliveros}]{kruparova2024unveiling}
Kruparova, O., Krupar, V., Szabo, A., {et~al.} 2024, The Astrophysical Journal Letters, 970, L13, \dodoi{10.3847/2041-8213/ad5da6}

\bibitem[{Kwak {et~al.}(2024)Kwak, Kim, Kim, Miyashita, Yang, Park, Lim, Jung, Kam, Lee, {et~al.}}]{kwak2024observational}
Kwak, Y.-S., Kim, J.-H., Kim, S., {et~al.} 2024, Journal of Astronomy and Space Sciences, 41, 171, \dodoi{10.5140/jass.2024.41.3.171}

\bibitem[{Lepping {et~al.}(1990)Lepping, Jones, \& Burlaga}]{lepping1990magnetic}
Lepping, R., Jones, J., \& Burlaga, L. 1990, Journal of Geophysical Research: Space Physics, 95, 11957, \dodoi{10.1029/ja095ia08p11957}

\bibitem[{Lepping {et~al.}(1995)Lepping, Ac{\~u}na, Burlaga, Farrell, Slavin, Schatten, Mariani, Ness, Neubauer, Whang, {et~al.}}]{lepping1995wind}
Lepping, R., Ac{\~u}na, M., Burlaga, L., {et~al.} 1995, Space Science Reviews, 71, 207

\bibitem[{Lin {et~al.}(1995)Lin, Anderson, Ashford, Carlson, Curtis, Ergun, Larson, McFadden, McCarthy, Parks, {et~al.}}]{lin1995three}
Lin, R., Anderson, K., Ashford, S., {et~al.} 1995, Space Science Reviews, 71, 125, \dodoi{10.1007/bf00751328}

\bibitem[{Manchester~IV {et~al.}(2005)Manchester~IV, Gombosi, De~Zeeuw, Sokolov, Roussev, Powell, K{\'o}ta, T{\'o}th, \& Zurbuchen}]{manchester2005coronal}
Manchester~IV, W., Gombosi, T., De~Zeeuw, D., {et~al.} 2005, The Astrophysical Journal, 622, 1225, \dodoi{10.1086/427768}

\bibitem[{Manchester~IV {et~al.}(2004)Manchester~IV, Gombosi, Roussev, De~Zeeuw, Sokolov, Powell, T{\'o}th, \& Opher}]{manchester2004three}
Manchester~IV, W.~B., Gombosi, T.~I., Roussev, I., {et~al.} 2004, Journal of Geophysical Research: Space Physics, 109, \dodoi{10.1029/2002JA009672}

\bibitem[{McComas {et~al.}(1998)McComas, Bame, Barker, Feldman, Phillips, Riley, \& Griffee}]{mccomas1998solar}
McComas, D., Bame, S., Barker, P., {et~al.} 1998, The advanced composition explorer mission, 563, \dodoi{10.1007/978-94-011-4762-0_20}

\bibitem[{Nakagawa(1993)}]{nakagawa1993solar}
Nakagawa, T. 1993, Solar physics, 147, 169, \dodoi{10.1007/bf00675493}

\bibitem[{Nakagawa {et~al.}(1989)Nakagawa, Nishida, \& Saito}]{nakagawa1989planar}
Nakagawa, T., Nishida, A., \& Saito, T. 1989, Journal of Geophysical Research: Space Physics, 94, 11761, \dodoi{10.1029/ja094ia09p11761}

\bibitem[{Neugebauer {et~al.}(1993)Neugebauer, Clay, \& Gosling}]{neugebauer1993origins}
Neugebauer, M., Clay, D., \& Gosling, J. 1993, Journal of Geophysical Research: Space Physics, 98, 9383, \dodoi{10.1029/93ja00216}

\bibitem[{Niembro {et~al.}(2019)Niembro, Lara, Gonz{\'a}lez, \& Cant{\'o}}]{niembro2019numerical}
Niembro, T., Lara, A., Gonz{\'a}lez, R.~F., \& Cant{\'o}, J. 2019, Journal of Space Weather and Space Climate, 9, A4, \dodoi{10.1051/swsc/2018049}

\bibitem[{Ohtani {et~al.}(2025)Ohtani, Zou, Merkin, Wiltberger, Pham, Raptis, Friel, \& Gjerloev}]{ohtani2025ground}
Ohtani, S., Zou, Y., Merkin, V., {et~al.} 2025, Journal of Geophysical Research: Space Physics, 130, e2024JA033691, \dodoi{10.1029/2024ja033691}

\bibitem[{Owens(2009)}]{owens2009formation}
Owens, M. 2009, Solar Physics, 260, 207, \dodoi{10.1007/s11207-009-9442-6}

\bibitem[{Owens {et~al.}(2017)Owens, Lockwood, \& Barnard}]{owens2017coronal}
Owens, M., Lockwood, M., \& Barnard, L. 2017, Scientific Reports, 7, 4152, \dodoi{10.1038/s41598-017-04546-3}

\bibitem[{Owens {et~al.}(2005)Owens, Cargill, Pagel, Siscoe, \& Crooker}]{owens2005characteristic}
Owens, M.~J., Cargill, P., Pagel, C., Siscoe, G., \& Crooker, N. 2005, Journal of Geophysical Research: Space Physics, 110, \dodoi{10.1029/2004ja010814}

\bibitem[{Pal {et~al.}(2020)Pal, Dash, \& Nandy}]{pal2020flux}
Pal, S., Dash, S., \& Nandy, D. 2020, Geophysical Research Letters, 47, e2019GL086372, \dodoi{10.1029/2019GL086372}

\bibitem[{Palmer {et~al.}(1978)Palmer, Allum, \& Singer}]{Palmer1978Bidirectional}
Palmer, I., Allum, F., \& Singer, S. 1978, Journal of Geophysical Research: Space Physics, 83, 75, \dodoi{10.1029/ja083ia01p00075}

\bibitem[{Palmerio {et~al.}(2016)Palmerio, Kilpua, \& Savani}]{palmerio2016planar}
Palmerio, E., Kilpua, E.~K., \& Savani, N.~P. 2016, Annales Geophysicae, \dodoi{10.5194/angeo-34-313-2016}

\bibitem[{Pollock {et~al.}(2016)Pollock, Moore, Jacques, Burch, Gliese, Saito, Omoto, Avanov, Barrie, Coffey, {et~al.}}]{pollock2016fast}
Pollock, C., Moore, T., Jacques, A., {et~al.} 2016, Space Science Reviews, 199, 331, \dodoi{10.1007/s11214-016-0245-4}

\bibitem[{Raghav {et~al.}(2023)Raghav, Shaikh, Vemareddy, Bhaskar, Dhamane, Ghag, Tari, Dayanandan, \& Mohammed Al~Suti}]{raghav2023possible}
Raghav, A., Shaikh, Z., Vemareddy, P., {et~al.} 2023, Solar Physics, 298, 64, \dodoi{10.1007/s11207-023-02157-y}

\bibitem[{Raghav {et~al.}(2019)Raghav, Choraghe, \& Shaikh}]{raghav2019cause}
Raghav, A.~N., Choraghe, K., \& Shaikh, Z.~I. 2019, Monthly Notices of the Royal Astronomical Society, 488, 910, \dodoi{10.1093/mnras/stz1608}

\bibitem[{Riley {et~al.}(2018)Riley, Baker, Liu, Verronen, Singer, \& G{\"u}del}]{riley2018extreme}
Riley, P., Baker, D., Liu, Y.~D., {et~al.} 2018, Space Science Reviews, 214, 1, \dodoi{10.1007/s11214-017-0456-3}

\bibitem[{Russell \& Shinde(2005)}]{russell2005defining}
Russell, C., \& Shinde, A. 2005, Solar Physics, 229, 323, \dodoi{10.1007/s11207-005-8777-x}

\bibitem[{Russell {et~al.}(2016)Russell, Anderson, Baumjohann, Bromund, Dearborn, Fischer, Le, Leinweber, Leneman, Magnes, {et~al.}}]{russell2016magnetospheric}
Russell, C., Anderson, B., Baumjohann, W., {et~al.} 2016, Space Science Reviews, 199, 189, \dodoi{10.1007/s11214-014-0057-3}

\bibitem[{Shaikh {et~al.}(2017)Shaikh, Raghav, \& Bhaskar}]{shaikh2017presence}
Shaikh, Z., Raghav, A., \& Bhaskar, A. 2017, The Astrophysical Journal, 844, 121, \dodoi{10.3847/1538-4357/aa729f}

\bibitem[{Shaikh {et~al.}(2018)Shaikh, Raghav, Vichare, Bhaskar, \& Mishra}]{shaikh2018identification}
Shaikh, Z.~I., Raghav, A.~N., Vichare, G., Bhaskar, A., \& Mishra, W. 2018, The Astrophysical Journal, 866, 118, \dodoi{10.3847/1538-4357/aae1b1}

\bibitem[{Smith {et~al.}(1998)Smith, L'Heureux, Ness, Acuna, Burlaga, \& Scheifele}]{smith1998ace}
Smith, C.~W., L'Heureux, J., Ness, N.~F., {et~al.} 1998, The advanced composition explorer mission, 613, \dodoi{10.1007/978-94-011-4762-0_21}

\bibitem[{Sonnerup \& Cahill(1967)}]{Sonnerup_1967}
Sonnerup, B. U.~Ã., \& Cahill, L.~J. 1967, Journal of Geophysical Research, 72, 171, \dodoi{10.1029/jz072i001p00171}

\bibitem[{St.~Cyr {et~al.}(2000)St.~Cyr, Howard, Sheeley~Jr, Plunkett, Michels, Paswaters, Koomen, Simnett, Thompson, Gurman, {et~al.}}]{st2000properties}
St.~Cyr, O., Howard, R., Sheeley~Jr, N., {et~al.} 2000, Journal of Geophysical Research: Space Physics, 105, 18169, \dodoi{10.1029/1999ja000381}

\bibitem[{Taylor(1974)}]{taylor1974relaxation}
Taylor, J.~B. 1974, Physical Review Letters, 33, 1139, \dodoi{10.1103/physrevlett.33.1139}

\bibitem[{Taylor(1986)}]{taylor1986relaxation}
---. 1986, Reviews of Modern Physics, 58, 741, \dodoi{10.1103/revmodphys.58.741}

\bibitem[{Teng {et~al.}(2024)Teng, Su, Ji, \& Zhang}]{teng2024unexpected}
Teng, W., Su, Y., Ji, H., \& Zhang, Q. 2024, Nature Communications, 15, 9198, \dodoi{10.1038/s41467-024-53538-1}

\bibitem[{Thampi {et~al.}(2025)Thampi, Bhaskar, Mayank, Vaidya, \& Venugopal}]{Thampi_2025}
Thampi, S.~V., Bhaskar, A., Mayank, P., Vaidya, B., \& Venugopal, I. 2025, The Astrophysical Journal, 981, 76, \dodoi{10.3847/1538-4357/ada93c}

\bibitem[{Tripathi {et~al.}(2022)Tripathi, Chakrabarty, Nandi, Raghvendra~Prasad, Ramaprakash, Shaji, Sankarasubramanian, Satheesh~Thampi, \& Yadav}]{Tripathi_Chakrabarty_Nandi_Raghvendra_Prasad_Ramaprakash_Shaji_Sankarasubramanian_Satheesh_Thampi_Yadav_2022}
Tripathi, D., Chakrabarty, D., Nandi, A., {et~al.} 2022, Proceedings of the International Astronomical Union, 18, 17–27, \dodoi{10.1017/S1743921323001230}

\bibitem[{Venugopal {et~al.}(2025)Venugopal, Thampi, \& Bhaskar}]{venugopal2025electrodynamic}
Venugopal, I., Thampi, S.~V., \& Bhaskar, A. 2025, Scientific Reports, 15, 26551, \dodoi{10.1038/s41598-025-08843-0}

\bibitem[{Vichare \& Bagiya(2024)}]{vichare2024manifestations}
Vichare, G., \& Bagiya, M.~S. 2024, Geophysical Research Letters, 51, e2024GL112569, \dodoi{10.1029/2024gl112569}

\bibitem[{Wang {et~al.}(2016)Wang, Zhang, Liu, Shen, Shen, Yang, Zic, Vrsnak, Webb, Liu, {et~al.}}]{wang2016propagation}
Wang, Y., Zhang, Q., Liu, J., {et~al.} 2016, Journal of Geophysical Research: Space Physics, 121, 7423, \dodoi{10.1002/2016ja022924}

\bibitem[{Webb \& Howard(2012)}]{webb2012coronal}
Webb, D.~F., \& Howard, T.~A. 2012, Living Reviews in Solar Physics, 9, 1

\bibitem[{Weiler {et~al.}(2025)Weiler, M{\"o}stl, Davies, Veronig, Amerstorfer, Amerstorfer, Le~Lou{\"e}dec, Bauer, Lugaz, Haberle, {et~al.}}]{weiler2025first}
Weiler, E., M{\"o}stl, C., Davies, E.~E., {et~al.} 2025, Space Weather, 23, e2024SW004260, \dodoi{10.1029/2024sw004260}

\bibitem[{Xie {et~al.}(2023)Xie, Gopalswamy, Akiyama, Yashiro, \& Makela}]{xie2023magnetic}
Xie, H., Gopalswamy, N., Akiyama, S., Yashiro, S., \& Makela, P. 2023, Journal of Atmospheric and Solar-Terrestrial Physics, 252, 106154

\bibitem[{Yadav {et~al.}(2025)Yadav, Vijaya, Krishnam~Prasad, Srikar, Mahajan, Mallikarjun, Zamani, B., Lohar, Kandpal, Narendra, Rai, Adoni, Veeresha, T.S., Kalpana, Nandita, \& Geeta}]{Yadav2025}
Yadav, V.~K., Vijaya, Y., Krishnam~Prasad, B., {et~al.} 2025, Solar Physics, 300, \dodoi{10.1007/s11207-025-02440-0}

\bibitem[{Zhang {et~al.}(2025)Zhang, Motoba, Paxton, \& Schaefer}]{zhang2025double}
Zhang, Y., Motoba, T., Paxton, L., \& Schaefer, R. 2025, AGU Advances, 6, e2025AV001688, \dodoi{10.1029/2025AV001688}

\end{thebibliography}
\bibliographystyle{aasjournal}

\end{document}